\documentstyle[aps]{revtex}
\begin{document}
\centerline{\LARGE\bf Interpreting the collapse process in terms
of} \centerline{\LARGE\bf new motion of particle} \vskip 1cm
\centerline{Gao shan} \centerline{Institute of Quantum mechanics}
\centerline{11-10, NO.10 Building, NO.21, YueTan XiJie DongLi,
XiCheng District} \centerline{Beijing 100045, P.R.China}
\centerline{E-mail: gaoshan.iqm@263.net}

\vskip 1cm
\begin{abstract}
Aiming at providing an objective picture for the collapse process
of wave function during measurement, further analysis about the
quantum discontinuous motion\cite{Gao} is presented, when
considering general relativity we show that the new motion is
essentially replaced by the quantum jump motion, which naturally
results in the collapse process of the wave function. Furthermore,
a concrete theoretical model is given to interpret the collapse
process quantitatively, and the coincidence between its
theoretical prediction and the experimental evidences is also
discussed. At last, the possibility to confirm the collapse model
is analyzed.
\end{abstract}

\vskip 1cm

\section{Introduction}
In order to provide a complete theory, the founders of quantum
mechanics cleverly added a simple projection postulate to its
consistent axiom system to account for the measurement
process\cite{Neumann}, whereas this postulate is evidently a
conditional description about the measurement process, it says
nothing about how the measurement can and does bring about one
definite result, or how the collapse process of the wave function
happens during measurement, so the projection postulate needs to
be further explained in
physics\cite{Anandan}\cite{Pea86}\cite{Pen86}.

In recent years many efforts have been made to tackle the
so-called collapse problem, and it is well known that in order to
bring about the collapse process of the wave function, some kind
of randomness should take effect in the normal deterministic
evolution of the wave function, but, up to now, the randomness has
not been introduced into on a solid basis yet, for example, in the
well-known GRWP
theory\cite{Ghir86}\cite{Ghir90}\cite{Pea86}\cite{Pea89} the
randomness comes from the special quantum fluctuations of
space-time, which is still ill-defined without a consistent
quantum theory of gravity; while in Penrose's proposal of
gravity-induced quantum state reduction\cite{Pen86}\cite{Pen96},
the origin of the randomness is not touched on at all, in fact,
based on the fundamental conflict between the principle of general
covariance of general relativity and the principle of
superposition of quantum mechanics, what he can conclude in a
heuristic way is only that the collapse process should happen when
relativistic gravity is considered in the evolution of the wave
function, since in the absence of randomness this conflict alone
can not provide an answer about how the collapse process happens.

Now, the new motion of particle\cite{Gao} just provides such a
broad framework for objectively studying the microscopic process
that it may also help to solve the collapse problem, for example,
there may exist many kinds of concrete motion modes among the new
motion, and the new motion may display differently in the
nonrelativistic and relativistic domains, especially when
involving relativistic gravity the new motion of particle may
naturally provide the origin of randomness in the collapse
process, and further result in the objective collapse process.

The plan of this paper is as follows: In Sect. 2 we first analyze
and give the natural origin of randomness in the collapse process.
Then in Sect. 3 the possible role of gravity in the collapse
process is carefully examined, and the proper existence of local
position state is demonstrated. In Sect. 4 we give a strict
definition of the quantum jump motion, and its evolution
principles are also presented, especially its relation with the
collapse process is deeply analyzed. In Sect. 5, as one example,
we analyze in detail the collapse process for two-state system
using the evolution principles of quantum jump motion, and the
concrete collapse time for such system is deduced out. In Sect. 6
we give a general evolution equation of the quantum jump motion,
which can be used to account for the collapse process of any wave
function. Then in Sect. 7 some experimental evidences supporting
our collapse model are discussed, and their coincidence with the
prediction of our collapse model is demonstrate. In sect. 8 we
give some further theoretical implications on the future theory of
quantum gravity. At last, in Sect. 9 we analyze the possibility to
confirm our collapse model using present technology.

\section{The natural origin of randomness in the collapse process}

As we have known, what the wave function objectively describes is
the new motion of the particle\cite{Gao}, while the essential
discontinuity of the new motion may provide the inherent
randomness in the collapse process from the inside. But in case of
no gravity this kind of randomness does not appear in the normal
deterministic evolution of the new motion state, since even during
infinitesimal time interval the particle undergoing the new motion
can still move throughout the whole space with a certain position
measure density $\rho(x,t)$, namely the essential discontinuity or
randomness is de facto absorbed in the definition of the new
motion state such as $\rho(x,t)$ etc; on the other hand, when the
normal deterministic evolution of the new motion is invalidated in
some situation, the veiled inherent randomness of the new motion
may appear, in the following we will analyze the possible
invalidation of normal deterministic evolution of the new motion.

According to the analysis about the new quantum discontinuous
motion\cite{Gao}, there are two important preconditions for the
normal deterministic linear evolution of the wave function, one is
the evolution of the wave function is studied in the
nonrelativistic domain, the other is the existence of the nonlocal
space-time reference framework.

As to the first precondition, when in nonrelativistic domain the
transfer of interaction is instantaneous, thus for the wave
function describing the quantum discontinuous motion of particle,
its different branches will not interact, this ensures the linear
superposition principle of the wave function at an exterior level,
when in relativistic domain the transfer velocity of interaction
is finite, thus for the wave function describing the quantum
discontinuous motion of particle, its different branches will
interact through the transfer particle of the interaction
including no gravity, for example, in quantum electrodynamics the
different branches of the electron wave function will interact
through the photon, which is characterized by the interaction term
$\bar \psi \gamma^{\mu} A_{\mu} \psi$, this interaction between
the different branches of the wave function will exteriorly
invalidate the linear superposition principle of the electron wave
function, but as we have done in relativistic quantum field
involving no gravity, this kind of interaction can still be
naturally and consistently included in the framework of the normal
linear evolution, so there exists no ultimate threat, and the
inherent randomness is still veiled in the state definition.

As to the second precondition, the existence of the nonlocal
space-time reference framework ensures the linear superposition
principle of the wave function at an interior level, since as we
have demonstrated\cite{Gao}, its existence will essentially result
in the existence of the equivalent nonlocal momentum description
of the new motion, which further determines the one-to-one
relation between the nonlocal momentum description and local
position description, then the most important linear superposition
principle of the wave function is deduced out, thus the existence
of the nonlocal space-time reference framework is the essential
basis of the linear superposition principle of the wave function,
or we can say, it ensures the normal deterministic linear
evolution of the wave function from the inside, and its
nonexistence may essentially result in the breakdown of the normal
deterministic linear evolution of the new motion, and the veiled
inherent randomness of the new motion may appear. In the
following, we will study where the nonlocal space-time reference
framework fails to exist, and how the veiled inherent randomness
of the new motion appears.

\section{The possible role of gravity in the collapse process}

\subsection{A general analysis}

As we have known, according to general relativity there does not
exist any nonlocal space-time reference framework at all, then the
usual nonlocal momentum eigenstate $e^{ipx}$ for describing the
normal new motion can not be properly defined, and the usual
linear superposition principle of the wave function will be broken
from the inside, thus the ubiquitous gravity will essentially
invalidate the above second precondition of the normal linear
evolution of the wave function.

Furthermore, in case of the nonexistence of any nonlocal
space-time reference framework, no nonlocal momentum state of the
new motion is permitted, and the local position state will be the
only proper state, thus the evolution of the new motion, as well
as the existence of the new motion, will be extremely different
from the normal one, although no complete formulation of quantum
gravity is in hand, we can still discuss the essential characters
of this kind of new existence and evolution involving relativistic
gravity, and give a crude but rational theoretical model to
account for the resulting collapse process, in the following
discussions we will always call relativistic gravity gravity for
simplicity.

\subsection{The existence of local position state}

On the one hand, when gravity comes into play, the only existence
of local position state for the new motion will essentially change
the motion mode of the new motion, concretely speaking, during
infinitesimal time interval the particle undergoing the new motion
can no longer move throughout the whole space with the original
position measure density $\rho(x,t)$, on the contrary, it can only
be in a local position state, we suppose the space size of this
local position state is $L_{g}$.

On the other hand, the discontinuous nature of the new motion
requires that during a finite time interval the particle
undergoing the new motion can still move throughout the whole
space with the original position measure density $\rho(x,t)$,
while in order not to destroy the local spreading law of energy
this finite time interval will be terrifically small, during which
the usual proper definition of energy, or even time, may not
exist, since this time interval provides a time limit of the valid
existence of the new motion state, we call this finite time
interval the minimal valid existing time $T_{e}$ of the new motion
state, or wave function describing the new motion, it further
means that if the particle stays in a local region much longer
than this time interval, then it will collapse into the local
state in this region, or we can say, this time interval may also
provide a determinant condition for the occurrence of the collapse
process; furthermore, the particle will generally stay in a local
region $L_{g}$ for a time interval much shorter than this time
interval, we assume the minimal average staying time in a local
region $L_{g}$ as $T_{g}$, and in the following we will discuss
how to find this minimal average staying time $T_{g}$ of the
particle.

\subsection{The implications of the space-time uncertainty}

As Karolyhazy et al have denoted\cite{Karolyhazy}, when combining
quantum mechanics and general relativity the uncertainty of
space-time is $\Delta T^{3}=T_{p}^{2}T$ and $\Delta
L^{3}=L_{p}^{2}L$, and Adler et al have recently re-demonstrated
that the well-known gravitional uncertainty principle $\Delta x
\geq \hbar/\Delta p+L_{p}^{2}\Delta p/\hbar$ is universally
valid\cite{Adler}, thus we assume that the above uncertainty
relation of space-time is an universal one, and we will further
analyze its deep implications.

First, when $T < T_{p}$ and $L
< L_{p}$ we have $\Delta T > T$ and $\Delta L > L$, then this
result evidently indicates that when $T < T_{p}$ and $L
< L_{p}$ the uncertainty
of time and space will be larger than the time and space
themselves, and the space-time stage on which everything is
defined will fall into ruins, thus the inherent uncertainty of
quantum mechanics makes itself invalid within the extraordinarily
small Planck time $T_{p}$ when satisfying general relativity, and
the definition of the wave function in quantum mechanics will be
also essentially invalid, or the usual nonlocal state of the
quantum discontinuous motion will be also ill-defined and
meaningless within the Planck size $T_{p}$ and $L_{p}$.

Secondly, the above uncertainty of space-time indicates that there
exists an absolute minimum uncertainty $L_{p}$ and $T_{p}$ for the
space-time where all particles exist and move, namely the minimum
distinguishable size of position and time of the particle are
$L_{p}$ and $T_{p}$, thus in physics the rational existence of the
particle is no longer in one position at one instant, but limited
in a space interval $L_{p}$ during a finite time interval $T_{p}$,
furthermore, during this finite time interval $T_{p}$ the particle
can only be limited in a space interval $L_{p}$, since if it can
move throughout at least two different local regions with
separation size larger than $L_{p}$ during the time interval
$T_{p}$, then there essentially exists a smaller distinguishable
finite time interval than $T_{p}$, which evidently contradicts the
above conclusion, thus the above uncertainty of space-time
essentially results in the rational existence of local position
state for the new motion, in which the particle stays in a local
region with size $L_{p}$ for a time interval $T_{p}$, and since
the time uncertainty formula $\Delta T^{3}=T_{p}^{2}T$ is
irrelevant to the mass of the particle, as well as the concrete
form of the wave function, this kind of local position state will
be the same for any wave function of any particle, and will be one
kind of general existence, we may call such general local position
state Planck cell state.

Thirdly, since the original new motion state or wave function in
quantum mechanics is defined during infinitesimal time interval,
the above conclusion will further result in that the proper state
which can be consistently defined should be only the local state,
and the new motion state or wave function can only be in use for a
finite time interval much larger than $T_{p}$, during which the
particle may move throughout the whole space with the position
measure density $\rho(x,t)=|\psi(x,t)|^{2}$ in quantum mechanics.

\section{The appearance of quantum jump motion}

"In essence, it is general relativity that turns the quantum
discontinuous motion into quantum jump motion."


Now, according to the above analysis, the quantum discontinuous
motion will be naturally replaced by a new kind of motion mode,
which is different from both quantum discontinuous motion and
classical continuous motion, we call it quantum jump motion.

\subsection{The definition of quantum jump motion}

Here we will give a strict definition about the quantum jump
motion.

(1). The quantum jump motion state of a particle in space is the
local position state, in which the particle stays in a local
region with size near $L_{p}$ for a time interval near $T_{p}$.

(2). During a time interval much larger than $T_{p}$, which is
still extremely smaller than usual time interval in quantum
mechanics, the particle will move throughout the whole space,
which average state is described by the position measure density
$\rho(x,t)=|\psi(x,t)|^{2}$ and position measure fluid density
$\mathbf{j}(x,t)$ as defined for the normal quantum discontinuous
motion.

(3). The evolution of the quantum jump motion is determined by
both quantum mechanics and general relativity.

The visual physical picture for the quantum jump motion will be
that during a finite time interval near $T_{p}$ the particle will
stay in a local region with size near $L_{p}$, then it will still
stay there or jump from this local region to another local region,
while during a time interval much larger than $T_{p}$ the particle
will move throughout the whole space with a certain position
measure density $\rho(x,t)$ as defined in quantum mechanics.

\subsection{A general discussion}

First, it is the requirement of general relativity that results in
that the quantum discontinuous motion is replaced by the quantum
jump motion, while the appearance of quantum jump motion just
releases the randomness inherent in the quantum discontinuous
motion, thus it will provide the objective origin of randomness or
probability in the collapse process.

Secondly, on the one hand, the particle undergoing the quantum
jump motion stays in a local region during infinitesimal time
interval, this is similar to the property of classical continuous
motion; on the other hand, during a finite time interval much
larger than $T_{p}$ the particle will continually jumps from one
local region to another local region, and move throughout the
whole space with a certain position measure density $\rho(x,t)$,
this is similar to the property of quantum discontinuous motion,
thus the quantum jump motion is evidently some kind of compromise
of quantum discontinuous motion and classical continuous motion,
and it will undoubtedly be responsible for the transition from
quantum discontinuous motion to classical continuous motion, or
from microscopic world to macroscopic world, at the same time, it
will be also the origin of the mysterious collapse process of the
wave function.

\subsection{The evolution principles of quantum jump motion}

As to the quantum jump motion, the particle does stay in a local
region for a finite nonzero time interval, and jump from this
local region to another local region stochastically, thus the
position measure density $\rho(x,t)$ of the particle will be
essentially changed by the finite nonzero stay time in a
stochastic way related to the stochastic stay region, and the
normal deterministic linear evolution of the wave function will be
also stochastically changed, in fact, as we will demonstrate, this
new element of stochastic evolution just plays an essential role
in generating the collapse process.

1. The first principle

As we have demonstrated, the quantum discontinuous motion is some
kind of average of the quantum jump motion during a finite time
interval much larger than $T_{p}$, thus the position measure
density $\rho(x,t)=|\psi(x,t)|^{2}$ of the particle will be also
the average of the position distribution of the particle
undergoing the quantum jump motion during this time interval, then
it is natural that the stochastic position in which the particle
stays will satisfy the position measure density $\rho(x,t)$, which
may evolve differently from the normal deterministic linear
evolution, namely the stochastic stay position of the particle
satisfies the distribution
\begin{equation}\label{}
P(x,t)=|\psi(x,t)|^{2}
\end{equation}
similar to the assumption in usual dynamical collapse
models\cite{Ghir86}\cite{Pea86}\cite{Dio87}\cite{Pea89}\cite{Dio89}\cite{Ghir90},
the probability of the stochastic stay position, which naturally
brings about the noise ad hoc introduced in usual dynamical
collapse models, is larger in the regions where the position
measure density $\rho(x,t)$ is larger, but here it is not an
implausible assumption, but a sound physical principle, which can
be naturally deduced out from the objective picture of quantum
jump motion.

2. The second principle

Now, according to the original definition of the position measure
density $\rho(x,t)$ for the quantum discontinuous motion, the
finite nonzero stay time for the quantum jump motion in a local
region evidently implies that the position measure density
$\rho(x,t)$ in that region will be increased after this finite
nonzero stay time interval, and the increase will be larger when
the stay time is longer.

We assume the particle undergoing the quantum jump motion stays in
a local region or Planck cell $L_{p}$ for a time interval $T_{p}$,
then according to the definition of the position measure density
its value $\rho_{c}(x,t)$ in this region will be increased, after
normalization it is
\begin{equation}\label{}
\rho_{c}(x,t)=\frac{\rho_{c}(x,t)+\Delta \rho}{1+\Delta \rho}
\end{equation}
where $\Delta \rho=T_{p}/\Delta t_{m}$, is the increase of the
position measure density $\rho_{c}(x,t)$ in this region, and $k$
is a dimensionless constant, $\Delta t_{m}\approx\hbar/\Delta E$,
is some average maximum permitted stay time, which evidently
results from the dimensional requirement and will be further
discussed in the following, in fact, this formula essentially
differs with the corresponding ad hoc assumption in usual
dynamical collapse models, since here it naturally results from
the existence of the objective quantum jump motion, which is
physically generated by the combination of general relativity and
quantum discontinuous motion.

Then we will further analyze the validity of the formula of
$\Delta \rho$, first, according to the original objective
definition of the position measure density $\rho(x,t)$ describing
the new motion, it essentially describes the real position measure
distribution of the particle during a time interval, and its
probability interpretation is only a derived exterior concept,
thus the change of the position measure density $\rho(x,t)$
resulting from the stay time will be essentially related to the
stay time itself, not its square or other forms, especially the
increase of the position measure density $\rho(x,t)$ will be
proportional to the stay time, when the stay time $\Delta t$ is
longer the increase of the position measure density $\rho(x,t)$ in
this stay region will be larger.

Secondly, we will analyze the validity of the concrete form of
$\Delta \rho$, and the origin and meaning of the average maximum
permitted stay time $\Delta t_{m}$, on the one hand, according to
the above general analysis, the existence of gravity will
essentially change the structure of space-time, and when combined
with the new quantum discontinuous motion this kind of change will
naturally result in the discrete quantum jump motion of the
particle with average minimum time interval $T_{p}$, thus it is
rational we assume the above time interval $T_{p}$ in the above
formula of $\Delta \rho$.

On the other hand, according to the principle of energy
conservation, during a finite nonzero time interval $\Delta t$ the
possible change of energy $\Delta E_{j}$ will be limited by the
uncertainty relation $\Delta E_{j} \approx \hbar / \Delta t$, then
the particle can hardly jump from this local region to another
local region when the difference of the whole energy $\Delta E$
between these two regions satisfies the condition $\Delta E \gg
\Delta E_{j}$,
namely after the stay time $\Delta t$ the position measure density
$\rho(x,t)$ will be greatly increased, concretely speaking, it
will be nearly one in this local region, and nearly zero in other
regions, in fact, the wave function has collapsed into this local
region in order to satisfy the requirement of energy conservation,
and this situation has also manifested that in order to satisfy
energy conservation the quantum jump motion will naturally result
in the collapse of the wave function; on the contrary, the
particle can more easily jump from this local region to another
local region when the difference of the whole energy $\Delta E$
between these two regions satisfies the condition $\Delta E \ll
\Delta E_{j}$, namely after the stay time $\Delta t$ the position
measure density $\rho(x,t)$ will be only changed slightly.

In one word, these two extreme situations have indicated that the
increase of the position measure density $\rho_{c}(x,t)$ will
relate to both the average minimal stay time $T_{p}$, which
denotes the limitation from the combination of general relativity
and quantum mechanics, and the average maximum permitted stay time
$\Delta t_{m}=\hbar/\Delta E$, or the difference of the whole
energy $\Delta E$ of the particle between the destination local
region and original region, which denotes the limitation from the
principle of energy conservation. Furthermore, the increase of the
position measure density $\rho(x,t)$ will be proportional to the
stay time $T_{p}$, and reversely proportional to the average
maximum permitted stay time $\Delta t_{m}=\hbar/\Delta E$, thus
when considering both the dimensional relation and first order
approximation, it is rational that we assume the relation $\Delta
\rho=k \Delta t/\Delta t_{m}$.


3. The third principle
In essence, we can not obtain the continuous limit of the above
evolution of the quantum jump motion in physics, since according
to the above analysis about the space-time uncertainty, there
exists a minimum stay time near $T_{p}$ of the particle undergoing
the quantum jump motion in reality, the motion is essentially
discrete, although when $T_{p} \rightarrow 0$ we can get the
continuous limit formally in mathematics.

\section{The collapse process for two-state system}

In this section, as one example we will first analyze the collapse
process for a simple two-state system, and work out the concrete
collapse time formula.

We suppose the initial wave function of the particle is
$\psi(x,0)=\alpha(0)^{1/2} \psi_{1}(x)+\beta(0)^{1/2}
\psi_{2}(x)$, which is a superposition of two static states with
different energy levels $E_{1}$ and $E_{2}$, which are located in
different regions $R_{1}$ and $R_{1}$ with the same spreading size
$L$, and we assume the relation $L \gg L_{p}$.

Now, for simplicity but lose no generality, we consider the space
of both static states as a whole local region, and only study the
quantum jump motion between these two regions, namely we directly
consider the difference of the whole energy $\Delta E=E_{2}-E_{1}$
between these two states, we assume the sequence of n jumps as
binary bit sequence
$[\delta_{n}]=\delta_{1}\delta_{2}...\delta_{k}...\delta_{n}$,
where $\delta_{k}=1$ denotes the particle stays in the region
$R_{1}$, $\delta_{k}=0$ denotes the particle stays in the region
$R_{2}$, the probability for this sequence as $P_{[\delta_{n}]}$,
then according to the above principles, if the particle stays in
the region $R_{1}$ for a time interval $T_{p}$ during the n-th
jump, the position measure density in this region will be changed
as follows:
\begin{equation}\label{}
\alpha_{[\delta_{n}]1}=\frac{\alpha_{[\delta_{n}]}+\Delta}{1+\Delta}
\end{equation}
\begin{equation}\label{}
\beta_{[\delta_{n}]1}=\frac{\beta_{[\delta_{n}]}}{1+\Delta }
\end{equation}
where $\Delta =k T_{p}/\Delta t_{m}$, and $\Delta
t_{m}=\hbar/\Delta E$; if the particle stays in the region $R_{2}$
for a time interval $T_{p}$ during the n-th jump, the position
measure density in this region will be changed as follows:
\begin{equation}\label{}
\alpha_{[\delta_{n}]0}=\frac{\alpha_{[\delta_{n}]}}{1+\Delta }
\end{equation}
\begin{equation}\label{}
\beta_{[\delta_{n}]0}=\frac{\beta_{[\delta_{n}]}+\Delta}{1+\Delta}
\end{equation}

Then similar to the continuous measurement theory for two-state
system, we can analyze the collapse process for the above
two-state system, first, we will calculate the diagonal elements
of the density matrix of the two-state system, according to the
definition of the quantum jump motion we have
\begin{equation}\label{}
\rho_{11}(n+1)=\sum_{[\delta_{n}]}(P_{[\delta_{n}]}
\alpha_{[\delta_{n}]}\alpha_{[\delta_{n}]1}+
P_{[\delta_{n}]}\beta_{[\delta_{n}]}\alpha_{[\delta_{n}]0})
\end{equation}
\begin{equation}\label{}
\rho_{22}(n+1)=\sum_{[\delta_{n}]}(P_{[\delta_{n}]}
\beta_{[\delta_{n}]}\beta_{[\delta_{n}]0}+
P_{[\delta_{n}]}\alpha_{[\delta_{n}]}\beta_{[\delta_{n}]1})
\end{equation}
while the definitions of the diagonal elements are
\begin{equation}\label{}
\rho_{11}(n)=\sum_{[\delta_{n}]}P_{[\delta_{n}]}\alpha_{[\delta_{n}]}
\end{equation}
\begin{equation}\label{}
\rho_{22}(n)=\sum_{[\delta_{n}]}P_{[\delta_{n}]}\beta_{[\delta_{n}]}
\end{equation}
and the initial conditions are
\begin{equation}\label{}
\rho_{11}(0)=\alpha(0)
\end{equation}
\begin{equation}\label{}
\rho_{22}(0)=\beta(0)
\end{equation}
then the discrete evolution equations will be
\begin{equation}\label{}
\rho_{11}(n+1)=\rho_{11}(n)
\end{equation}
\begin{equation}\label{}
\rho_{22}(n+1)=\rho_{22}(n)
\end{equation}
when in the continuous mathematical limit, these evolution
equations for the diagonal elements will be
\begin{equation}\label{}
\dot{\rho}_{11}(t)=0
\end{equation}
\begin{equation}\label{}
\dot{\rho}_{22}(t)=0
\end{equation}

Now it is evident that the above principles essentially guarantee
that the collapse results distribution satisfies the position
measure density $\rho(x,t)$, this is consistent with the
prediction of quantum mechanics.

Secondly, in order to work out the concrete collapse time for the
above two-state system, we will calculate the non-diagonal
elements of the density matrix, similar to the above calculations
we can get the discrete evolution equation, namely
\begin{equation}\label{}
\rho_{12}(n+1)=[1-(\frac{\Delta}{1+\Delta})^{2}]^{1/2}\rho_{11}(n)
\end{equation}
then the solution will be
\begin{equation}\label{}
\rho_{12}(n)=[1-(\frac{\Delta}{1+\Delta})^{2}]^{n/2}\rho_{12}(0)
\approx(1-\frac{1}{2}\Delta^{2}n)\rho_{12}(0)
\end{equation}
and the collapse formula is
\begin{equation}\label{}
\nu_c=1-\frac{1}{2}\Delta ^{2}n
\end{equation}
Now according to the relation $\Delta =k T_{p}/\Delta t_{m}=k
\Delta E/E_{P}$, where $E_{P}=\hbar/T_{p}$, is the Planck energy,
$T_{p}$ is the average minimum stay time of the particle in one of
the two local regions during each jump, and the relation
$n=\tau/T_{p}$, where $\tau$ is the whole time interval of the
quantum jump motion, the above collapse formula will be
\begin{equation}\label{}
\nu_c=1-\tau/\tau_{c}
\end{equation}
where $\tau_{c} \approx \frac{2}{k^{2}}\cdot\frac{E_{p}}{\Delta
E}\cdot\frac{\hbar}{\Delta E}$, is the collapse time.

Then we will further discuss the collapse time formula, first,
according to the above general discussions about the collapse
process, when $\Delta E=E_{p}$, the collapse time will be $T_{p}$,
then we may simply assume $k=\sqrt{2}$, and the collapse time will
be $\tau_{c}=\frac{\hbar}{\Delta E}$, this evidently denotes that
the normal quantum evolution of the wave function is invalid at
the critical point $\Delta E\approx E_{p}$, and in fact the
evolution is mainly determined by the collapse process, thus we
may call the Planck energy $E_{p}$ collapse critical energy; when
$\Delta E \ll E_{p}$, namely for microscopic objects, we get
$\tau_{c} \gg \frac{\hbar}{\Delta E}$, this result means that for
microscopic objects the collapse time is much longer than the
normal evolution time of the wave function, thus quantum mechanics
is still approximately valid before the collapse process happens;
while when $\Delta E \gg E_{p}$, namely for macroscopic objects,
we get $\tau_{c} \ll \frac{\hbar}{\Delta E}$, which means that for
macroscopic objects the collapse time is much shorter than the
normal evolution time of the wave function in quantum mechanics,
then quantum mechanics will be naturally replaced by classical
mechanics. In fact, the border between microscopic objects and
macroscopic objects mostly results from the experience and sense
of our mankind, thus the above discussion may be improper, the
relevant criterion may be that if the collapse time of the objects
is small in our sense, say a small section of one second, we may
call such objects macroscopic objects, then for macroscopic
objects we may assume $\Delta E \geq 10Mev$.

Secondly, since the above difference of energy $\Delta E$ is
irrelevant to the position measure density $\rho(x,t)$ of the wave
function, the above collapse time is also irrelevant to the
initial position measure density $\rho(x,t)$ of the wave function,
in fact, this conclusion essentially results from the existence of
quantum jump motion and assumed dynamical continuity of the
collapse process.

Thirdly, it can be easily seen that the above collapse formula is
just Fivel's assumed formula resulting from the experimental
considerations\cite{Fivel}, but the meaning of $\Delta E$ is a
little different, here the difference of energy $\Delta E$ will
include all the energy difference obtained from the last unified
theory, especially include the possible difference of the gravity
energy of the particle in the states, which will essentially
result from the complete theory of quantum gravity, while the
difference of energy in Fivel's formula is just the dispersion of
the usual energy of the particle in quantum mechanics, the obvious
distinction of these two kinds of definitions lies in that the
former may relate to the space separation of the wave function,
while the latter relates not to the space separation of the wave
function, although in general situation the difference of the
gravity energy of the particle may be proportional to the
dispersion of the usual energy of the particle.

Fourthly, the existence of the above collapse formula will imply
the nonexistence of the continuous limit in physics, namely in
reality there should exists a nonzero minimum average stay time
$T_{p}$ for the particle undergoing the quantum jump motion, and
this minimum average stay time will also appear in the collapse
formula, and determine the collapse time of the wave function,
this also results in the appearance of the collapse critical
energy scale $E_{p}$ in the above collapse formula, in the
following, we will further analyze the validity of the formula of
the collapse critical energy $E_{p}=\sqrt{\hbar c^{3}/G}$. First,
when the new quantum discontinuous motion does not exist, namely
$\hbar=0$, this energy scale $E_{p}$ will turn to be zero, since
the continuous motion has been local, it needs no additional
energy to collapse into a local region; secondly, when gravity
disappears, namely $G=0$, this energy scale $E_{p}$ will turn to
be infinite, since the collapse process never happens; thirdly,
when in nonrelativistic domain, namely $c\rightarrow \infty$, the
collapse process will also never happen, then it is also rational
that when $c\rightarrow \infty$ this energy scale $E_{p}$ turns to
be infinite, thus from the analysis about the constants involved
in $E_{p}$ the formula of the collapse critical energy
$E_{p}=\sqrt{\hbar c^{3}/G}$ is rational, and will be the only one
when involving only the above constants in the formula.
Furthermore, this conclusion also strongly implies that the Planck
energy $E_{p}$ will appear in the collapse time formula, thus
Penrose's formula of gravity-induced quantum state reduction may
be inappropriate, while the above collapse formula will be more
appropriate.

At last, the problem of energy conservation needs to be
considered, as Fivel has pointed out\cite{Fivel}, the collapse
formula $\tau_{c} \approx \frac{E_{p}}{\Delta
E}\cdot\frac{\hbar}{\Delta E}$ guarantees the principle of energy
conservation, now the above collapse model resulting from the
continual jump of the particle undergoing the quantum jump motion
will be also consistent with the principle of energy conservation,
concretely speaking, for any single particle in the state, the
energy change during the collapse process lies within the limit of
the uncertainty principle, while for the ensemble of the particles
in the state the average energy change is evidently zero, since
the collapse result is one of the energy eigenstates, and the
distribution of the collapse results is just the initial
distribution of the energy eigenstates.

Furthermore, we will demonstrate that for a single particle in the
state, the energy change during the collapse process lies within
the limit of the uncertainty principle, first, during any single
jump the collapse critical energy scale $E_{p}$( or the maximum
permitted energy change ) and the minimal stay time interval
$T_{p}$ satisfies the energy uncertainty relation
$E_{p}=\hbar/T_{p}$, and the principle of energy conservation is
naturally satisfied; secondly, when considering the whole collapse
process Fivel has demonstrated that the principle of energy
conservation is also satisfied\cite{Fivel}, here we give some
further discussions, since when $\Delta E<E_{p}$, we have the
relation $\tau_{c} \approx \frac{E_{p}}{\Delta
E}\cdot\frac{\hbar}{\Delta E}>\frac{\hbar}{\Delta E}$, this means
that the energy change during the collapse process lies within the
limit of the uncertainty principle; while when $\Delta E \geq
E_{p}$, we have $\tau_{c}  \approx \frac{E_{p}}{\Delta
E}\cdot\frac{\hbar}{\Delta E}\leq \frac{\hbar}{\Delta E}\leq
T_{p}$, this relation implies that the collapse process happens
within the minimum stay time interval $T_{p}$, namely such
superposition states with the energy difference larger than
$E_{p}$ can not exist at all, thus no problem of energy
conservation is involved for this situation. In one word, the
above collapse model and the resulting collapse formula $\tau_{c}
\approx \frac{E_{p}}{\Delta E}\cdot\frac{\hbar}{\Delta E}$ is
consistent with the principle of energy conservation.

\section{The general evolution equation of the quantum jump
motion}

In the following, we will give a general evolution equation of the
quantum jump motion, which can be used to account for the collapse
process of any wave function describing the quantum jump motion.

For simplicity but lose no generality, we consider a one-dimension
initial wave function $\psi(x)$, whose spreading size is $L$, and
divide the whole size of the wave function $L$ into many local
Planck cells with size $L_{p}$, which are denoted by number $i$,
the basic local position state of the particle is limited in one
of such local cells, then according to the above analysis, the
concrete evolution equation of the quantum jump motion will be
essentially one kind of revised stochastic evolution equation
based on Schr\"{o}dinger equation for the wave function, here we
adopt the stochastic differential equation ( SDE ), it is
\begin{equation}\label{}
d\psi(x,t)=\frac{1}{i\hbar}H_{Q}\psi(x,t)dt+\frac{1}{2}[\frac{\delta
(x-x_{N})}{\rho(x,t)}-1] \frac{\Delta
E(x_{N},\overline{x_{N}})}{E_{p}}\psi(x,t)dt
\end{equation}
where $\delta(x-x_{N})$ is the discrete $\delta$-function,
$\rho(x,t)=|\psi(x,t)|^{2}$, is the measure density, $\Delta
E(x_{N},\overline{x_{N}})$ is the total difference of the energy
of the particle between the cell containing $x_{N}$ and all other
cells $\overline{x_{N}}$, $x_{N}$ is a stochastic position
variable, whose distribution is
$P(x_{N},t)=\rho(x_{N},t)=|\psi(x,t)|^{2}$. In physics, this
stochastic differential equation should be taken as a continuous
mathematical description of the discrete evolution equation of the
quantum jump motion.

\section{Some experimental evidences}
Even though we have not directly confirmed the above collapse
model up to now, there still exists some experimental evidences
which may support the above collapse model, in fact, these
coincidences have strongly implied its validity, and confuted
other collapse models.

The first experimental evidence is the gravity-induced CP
violations discussed by Fivel, as he analyzed in an elegant
paper\cite{Fivel}, the observed magnitude of CP violation in
$K_{L}$ meson decay remarkably indicates that the collapse process
may happen during the decay and the collapse time satisfies the
above collapse formula when we assume $k\approx 4\sqrt{\pi}$,
while according to Penrose's formula, the collapse time will be
infinite, since the difference of the gravity self-energy of the
states in the superposition is zero, but this result can hardly
account for the bizarre coincidence of the observed magnitude of
CP violation with the above prediction of collapse time; at the
same time, the above collapse model may also provide an economical
explanation of the origin of the CP violations, namely all the CP
violations just results from the gravity-induced collapse of the
wave function of the particle, which has been also demonstrated
from other perspects\cite{Ahlu}.

The second experimental evidence comes from the avalanche
photodiodes devices used for detecting the particles in short time
intervals, just as Berg has analyzed\cite{Berg}, even if the time
resolutions and energy consumption of these devices can not
translate into an immediate estimate of the collapse time, we can
still estimate the upper bounds of the collapse time, his
calculation indicates that $b=2E_{p}/k^{2}\Delta
E\approx10^{17}<3.8\times 10^{21}$, namely the prediction of the
above collapse formula is consistent with the detection property
of these measuring devices; on the other hand, according to
Penrose's formula, the collapse time will be
$\tau_{c}=b\frac{\hbar}{\Delta E} \approx \frac{\hbar}{\Delta
E_{g}}$, and the equivalent value of $b$ is $b\approx10^{40} \gg
3.8\times 10^{21}$, thus this result has evidently invalidated
Penrose's formula.

The third experimental evidence comes from the experimental
results about the Ramsey fringes in atomic beam spectroscopy,
according to Berg's calculations\cite{Berg}, we get the evident
relation $b=2E_{p}/k^{2}\Delta E\approx10^{28}>1.35\times
10^{11}$, which indicates that the experiments in atomic beam
spectroscopy are also consistent with the above collapse formula.

\section{Some further theoretical implications}
Although it may be very difficult to formulate a complete theory
of quantum gravity on the basis of the quantum jump motion and
resulting collapse process, we can still discuss its direct
implications in theory, which will be very valuable for
understanding the nature of quantum gravity.

First, the existence of the quantum jump motion naturally results
from the combination of quantum mechanics and general relativity,
and consistent with the well-known gravitional uncertainty
relation\cite{Adler}, according to which the continuous space-time
concept can not be properly used within the Planck scale $T_{p}$
and $L_{p}$, the local position state will be the only proper
state for the quantum motion, in fact, this conclusion is also
implied by the fact that a particle with mass larger than $M_p =
E_p/c^2 \approx 10^{-5}$gm has a Compton wavelength smaller than
its Schwarzschild radius.

Secondly, as Feynman suggested in his {\em Lectures on
Gravitation}\cite{Feynman}, the quantum jump motion just provides
the objective reason that gravity need not be quantized, because
when involving gravity the linear Schr\"{o}dinger evolution of the
wave function will be broken, and replaced by the nonlinear
stochastic evolution, which, as we have demonstrated, will result
in the immediate collapse of the wave function at the Planck scale
$E_p$.

Thirdly, since the Planck mass $M_p$ is essentially macroscopic,
the above collapse model based on the quantum jump motion can
easily explain the apparent instability and collapse of
Schr\"{o}dinger cat states in which there is dispersion of a
macroscopic observable, and will be responsible for the transition
from the microscopic world to the macroscopic world, furthermore,
the evolution of the quantum jump motion also naturally provides a
generalization of the Schr\"{o}dinger equation that interpolates
between linear evolution and collapse, which may unify the
descriptions about the microscopic world and macroscopic world.

Fourthly, as to the above collapse model based on the objective
quantum jump motion, there does not exist the usual tails problem
inherent in other dynamical collapse models at
all\cite{Ghir86}\cite{Pea86}\cite{Dio87}\cite{Pea89}\cite{Dio89}\cite{Ghir90},
and arithmetic can still apply to ordinary macroscopic
objects\cite{Lewis}, since according to our collapse model, the
collapse process results from the quantum jump motion, which is
essentially discrete, the continuous stochastic evolution of the
wave function is just an illusion, especially in the last stage of
the collapse process, when the particle stays in one of the
branches long enough it will de facto collapse into that branch
owing to the limitation of energy conservation, and the ostensible
description---wave function also completely disappears in other
branches, while this will never happen if the continuous wave
function is taken as the last objective description of the quantum
process, and the tails will survive forever.

At last, the existence of the quantum jump motion will naturally
tackle the well-known time problem involved in formulating a
complete theory of quantum gravity\cite{Pen96}\cite{Isham}, since
as to the quantum jump motion, the local position state will be
the only proper state, and during a finite time interval near
$T_{p}$, let alone at one instant, the particle can only be
limited in a space interval near $L_{p}$, namely there does not
exist any essential superposition of different positions at all,
thus the essential inconsistency of the superposition of different
space-time in the theory of quantum gravity will naturally
disappear, and the new physical picture based on the quantum jump
motion will be that at any instant the structure of the space-time
determined by the ostensible superposition state in quantum
mechanics is definite or "classical", while during a finite time
interval it will be stochastically disturbed by the discontinuous
quantum jump motion of the particle in the superposition state,
and this stochastic disturbance picture just reflects the real
quantum nature of the space-time and matter.

Certainly, the essential nonexistence of the superposition of
different space-time structure will also invalidate Penrose's
proposal of gravity-induced quantum state
reduction\cite{Pen86}\cite{Pen96}, even though his start point
involving gravity is undoubtedly right, since as to the quantum
jump motion, there does not exist any essential "fuzziness" of the
time translation, in fact, it has been replaced by the stochastic
disturbance of space-time resulting from the quantum jump motion
of the particle, thus his collapse formula $\tau_{c} \approx
\frac{\hbar}{\Delta E}$ will be essentially improper.

\section{Some further considerations about the experimental confirmation}
Although some experiments have been presented to confirm the usual
dynamical collapse models and gravity-induced collapse models, it
may be very difficult to implement them using present technology,
and many efforts need to be made along this direction, here we
will further discuss some possible experiments to confirm our
collapse model.

At first, we will analyze the general relation between the above
collapse process and environment-induced decoherence, since the
decoherence process will generally introduce the energy difference
among the corresponding decoherence branches of the wave function,
and further result in the dynamical collapse process according to
our collapse model, at the same time, the decoherence effects will
also prevent us from detecting the collapse effects, these two
processes are closely interconnected, through a simple
mathematical analysis we can find the general relation between the
collapse time $\tau_{c}$ and decoherence time $\tau_{d}$, it is
\begin{equation}\label{}
\tau_{c}=(\frac{\hbar
E_{p}}{k^{2}})^{\frac{1}{3}}(\frac{\tau_{d}}{\gamma
T})^{\frac{2}{3}}\approx10^{7}(\frac{\tau_{d}}{\gamma
T})^{\frac{2}{3}}
\end{equation}
where $k$ is Boltzmann constant,
$\gamma=4\varepsilon/(1+\varepsilon)^{2}$, is the energy transfer
factor during a single interaction with the particles in the
environment, $\varepsilon=E_{i}/E$ is the energy ratio, $E_{i}$ is
the energy of the particles in the environment, $T$ is the
temperature of the environment.

From this relation between the collapse time $\tau_{c}$ and
decoherence time $\tau_{d}$, we can see that the decoherence
process will generally happen before the collapse process, thus
for this situation we can not detect and confirm the existence of
the collapse process; on the other hand, if the temperature of the
environment is so high, say when $T>10^{10}$, the collapse process
will happen before the decoherence process and can be detected,
thus this situation will provide the possible experimental
confirmation of our collapse model.

Secondly, we will further analyze the above possibility to confirm
our collapse model, on the one hand, it may exist everywhere in
the universe, say inside the stars, on the other hand, we had
better devise a similar experiment in the earth, as we know, for
this situation, in a single interaction the energy transfer will
be very large, while the dephasing effect is smaller than the
collapse effect resulting from the increased difference of energy,
here we consider a trapped neural particle, say a neuron, in a
initial superposition state
$\psi=\frac{1}{\sqrt{2}}(\psi_{1}^{i}+\psi_{2}^{i})$ of different
positions with the same energy, and let a high-energy neural atom
in state $\varphi$ with energy $E$ collide with one of the neuron
states, say $\psi_{1}$, then the state of the whole system after
interaction will be
$\omega=\frac{1}{\sqrt{2}}(\psi_{1}^{o}\varphi_{1}+\psi_{2}^{o}\varphi_{2})$,
then the energy difference between the neuron states
$\psi_{2}^{o}$ and $\psi_{2}^{o}$ will be approximately
$4m_{n}E/m_{a}$, and the whole energy difference in the state
$\omega$ will be $8m_{n}E/m_{a}$, thus according to our collapse
model, this state will collapse into the branch
$\psi_{1}^{o}\varphi_{1})$ or $\psi_{2}^{o}\varphi_{2})$ after a
collapse time
\begin{equation}\label{}
\tau_{c}=(\frac{m_{a}}{8m_{n}})^{2}\frac{\hbar E_{p}}{E^{2}}
=\frac{1}{16}\frac{\hbar E_{p}}{m_{n}^{2}v^{4}}
\end{equation}
where $v$ is the velocity of the collision particle, while the
dephasing effect of this process will be very small, since the
mass of the atom is assumed to be very larger than that of the
neutron, and the difference between the states of the atom
$\varphi_{1}$ and $\varphi_{2}$ will be very small, and not change
with time, namely $<\varphi_{1}|\varphi_{2}>\approx
1-4m_{e}/m_{n}$, at the same time, we assume the decoherence
effect resulting from the gravitational field can also be omitted,
thus one of the direct ways is to keep the last state $\omega$
long enough, then we can easily detect the predicted collapse
effect, the concrete relation will be
$t>\frac{m_{n}}{16m_{e}}\frac{\hbar E_{p}}{E^{2}}$.

From the above relation we can see that, there exists an optimal
mass for the collision particle limited by present technology, if
its mass is too large, we can hardly generate higher velocity,
then the collapse time will be too long and it will be very
difficult to keep the last state; while if its mass is too small,
its energy $E$ will be also too small, and the collapse time will
be also too long.

Now, as one example, we let $m_{n}/m_{a}=100$ and
$E=E_{n}/100=9.3Mev$, which means that the velocity of the atom is
approximately $c/100$, then we get the collapse time
$\tau_{c}\approx10^{4}s$, which is in the level of minutes, so if
we can keep the last state of the whole system longer than
$4m_{n}/m_{a}\tau_{c}\approx 100s$, we can differentiate the
dephasing effect and collapse effect, and confirm the existence of
the assumed collapse process.

\section{Conclusions}
On the whole, we present a concrete collapse model on the basis of
the new quantum discontinuous motion, according to this model, it
is the combination of general relativity and quantum mechanics
that turns the quantum discontinuous motion into the quantum jump
motion, and further results in the collapse process of the wave
function, while the resulting quantum jump motion just provides
the inherent randomness or probability involved in the collapse
process. Furthermore, a crude theoretical model is given to
interpret the collapse process quantitatively, and the consistence
between theoretical prediction and experimental evidences is also
discussed. At last, we analyze the possibility to confirm the
collapse model using present technology.

\vskip .5cm \noindent Acknowledgments \vskip .5cm Thanks for
helpful discussions with X.Y.Huang ( Peking University ),
A.Jadczyk ( University of Wroclaw ), P.Pearle ( Hamilton College
), F.Selleri ( University di Bari ), Y.Shi ( University of
Cambridge ), A.Shimony, A.Suarez ( Center for Quantum Philosophy
), L.A.Wu ( Institute Of Physics, Academia Sinica ), Dr S.X.Yu (
Institute Of Theoretical Physics, Academia Sinica ), H.D.Zeh.

\end{document}